\documentstyle[emulateapjdon,flushrt,psfig]{article}
\received{9 May 2000}
             \font\sevenrm=cmr7

\def\mathrm#1{{\hbox{\sevenrm #1}}}

\newcommand\Emax{E_{\rm max}}

\def\xx#1{\!\times\!10^{#1}}

{\catcode`\@=11
\gdef\SchlangeUnter#1#2{\lower2pt\vbox{\baselineskip 0pt\lineskip0pt
\ialign{$\m@th#1\hfil##\hfil$\crcr#2\crcr\sim\crcr}}}}
\def\gtrsim{\mathrel{\mathpalette\SchlangeUnter>}}
\def\lesssim{\mathrel{\mathpalette\SchlangeUnter<}}
\def\itt{ }
\def\bff{ }
\def\aaDCE#1#2#3{ 19#1, A\&A, #2, #3}
\def\aasupDCE#1#2#3{ 19#1, {\itt A\&AS,} {\bff #2}, #3}
\def\ajDCE#1#2#3{ 19#1, {\itt AJ,} {\bff #2}, #3}

\def\apjDCE#1#2#3{ 19#1, {\itt ApJ,} {\bff #2}, #3}
\def\apjletDCE#1#2#3{ 19#1, {\itt ApJ(Letts),} {\bff  #2}, #3}

\def\apjletpress{{\itt ApJ(Letts),} in press}
\def\apjsDCE#1#2#3{ 19#1, {\itt ApJS,} {\bff #2}, #3}

\def\apjsub2DCE#1{ 20#1, {\itt ApJ}, submitted.}

\def\assDCE#1#2#3{ 19#1, {\itt Astr. Sp. Sci.,} {\bff #2}, #3}
\def\icrcsaltlake#1#2{ 1999, {\itt Proc. 26th Int. Cosmic Ray Conf.
    (Salt Lake City),} {\bff #1}, #2.}

\def\jgrDCE#1#2#3{ 19#1, {\itt J.G.R., } {\bff #2}, #3}
\def\mnrasDCE#1#2#3{ 19#1, {\itt M.N.R.A.S.,} {\bff #2}, #3}
\def\natureDCE#1#2#3{ 19#1, {\itt Nature,} {\bff #2}, #3}
\def\phyreptsDCE#1#2#3{ 19#1, {\itt Phys. Repts.,} {\bff #2}, #3}

\def\rppDCE#1#2#3{ 19#1, {\itt Rep. Prog. Phys.,} {\bff #2}, #3}
\def\ssrDCE#1#2#3{ 19#1, {\itt Space Sci. Rev.,} {\bff #2}, #3}
\def\TP{test-particle}
\def\NL{nonlinear}

\def\etainjP{\eta_\mathrm{inj,p}}

\def\muG{$\mu$G}

\def\iec{i.e., }
\def\egc{e.g., }
\def\etal{et al.}

\def\alf{Alfv\'en}

\slugcomment{Accepted in The Astrophysical Journal Letters}

\begin{document}

\newcommand{\vol}[2]{$\,$\rm #1\rm , #2.}

\vphantom{p}
\vskip -40pt 
\vskip 15pt
%


\title{Thermal X-ray Emission and Cosmic Ray
Production in Young Supernova Remnants}

\author{
Anne ~Decourchelle\altaffilmark{1},
Donald C.~Ellison\altaffilmark{2}, and
Jean Ballet\altaffilmark{1}
}
\altaffiltext{1}{Service d'Astrophysique, L'Orme des Merisiers,
CE-Saclay, 91191 Gif-sur-Yvette, CEDEX France;
E-mail: adecourchelle@cea.fr, jballet@cea.fr}

\altaffiltext{2}{Department of Physics, North Carolina State University,
 Box 8202, Raleigh NC 27695, U.S.A.;
E-mail: don\_ellison@ncsu.edu}

\begin{abstract}
We have developed a simple model to investigate the
modifications of the hydrodynamics and non-equilibrium ionization X-ray
emission in young supernova remnants due to nonlinear particle
acceleration.
In nonlinear, diffusive shock acceleration, the heating of the gas to
X-ray emitting temperatures is strongly coupled to the acceleration of
cosmic ray ions. If the acceleration is efficient and a significant fraction
of the shock ram energy ends up in cosmic rays, compression ratios will be
higher and the shocked temperature lower than test-particle,
Rankine-Hugoniot
relations predict.
We illustrate how particle acceleration impacts the interpretation of X-ray
data using the X-ray spectra of Kepler's remnant, observed by {\it ASCA} and
{\it RXTE}. The thermal X-ray emission provides important constraints on
the efficiency of particle acceleration, in complement to nonthermal
emission. X-ray data from {\it Chandra} and {\it XMM Newton}, plus radio
observations, will be essential to quantify nonlinear effects.
\end{abstract}

\keywords{Supernova remnants --- acceleration of particles --- hydrodynamics
--- X-rays: ISM --- ISM: individual (Kepler) --- ISM: abundances}

\section{Introduction}

The strong forward and reverse shocks in young supernova remnants (SNRs)
heat the ambient interstellar medium (ISM) and supernova ejecta material to
X-ray emitting temperatures.  This X-ray emission contains information on
the supernova (e.g., type, explosion energy, mass and composition of
ejecta) and on the nature of the ambient medium (e.g., uniform, stellar
wind).

In addition to heating the plasma, the forward and reverse shocks are
thought to accelerate some fraction of the shocked material to cosmic ray
energies. A large body of evidence suggests that this acceleration
can be quite efficient with as much as 50\% of the ram kinetic energy
of the ejecta being placed in relativistic particles and removed from
the thermal plasma (\egc Blandford \& Eichler \cite{BE87}; Jones \& Ellison
\cite{JE91}; Kang \& Jones \cite{KJ91}; Berezhko, Ksenofontov, \& Petukhov
\cite{BereKP99}), and some early work investigated its observational
impact (Heavens \cite{Heavens84}, Boulares \& Cox \cite{Boucox88}, Dorfi \&
B\"ohringer \cite{DorfiB93}). However, despite the expected efficiency of
shock acceleration, the modeling and interpretation of X-ray observation
from SNRs are generally done assuming that the shocks {\it do not} place a
significant fraction of their energy in cosmic rays.

The recent discovery of nonthermal X-ray emission in shell-like SNRs like
SN1006 (Koyama et al. \cite{Koyama95}), Cas A (Allen et al. \cite{Allen97}),
G347.3-0.5 (Slane et al. \cite{Slane99}) and RCW86 (Borkowski \etal\
\cite{Borko00}) has shown that SNR shocks do accelerate electrons to TeV
energies and supports the case for efficient proton acceleration and the
necessity of investigating its effects in the modeling of thermal X-ray
emission.

Besides providing a more self-consistent SNR model, X-ray models of
cosmic ray modified SNRs provide a test of the fundamental assumption that
SNRs are the primary sources of galactic cosmic ray ions. Since shocks put
more energy into accelerated ions than electrons, \NL\ effects seen in X-ray
emission will be evidence for the efficient shock acceleration of ions as
well.

Here, we investigate the effects of efficient particle acceleration on the
hydrodynamics and thermal X-ray emission of young SNRs.
We have developed a model which couples self-similar hydrodynamics
(Chevalier \cite{Chev83}) to \NL\ diffusive shock acceleration
(Berezhko \& Ellison \cite{BEapj99}; Berezhko \etal\ \cite{BereKP99};
Ellison, Berezhko, \& Baring \cite{EBB2000}), and X-ray emission including
non-equilibrium ionization effects (Decourchelle \& Ballet
\cite{Decour94}). We show that a modified self-similar approach is relevant
in most cases and compare it to the standard unmodified \TP\ case.

\section{Nonlinear Shock Acceleration}

There is convincing direct and indirect evidence from spacecraft
observations and plasma simulations that collisionless shocks are
efficient \NL\ accelerators.  Consequently, they self-generate
magnetic turbulence and develop a smooth shock precursor
(\egc Drury \cite{Drury83}; Blandford \& Eichler \cite{BE87}; Jones \&
Ellison \cite{JE91}; Achterberg, Blandford, \& Reynolds \cite{ABR94};
Terasawa \etal\ \cite{Terasawa99}; Ellison \etal\ \cite{EBB2000}).
When the acceleration is \NL, the shock compression ratio
increases and the shocked temperature decreases compared to \TP\ values.
The entire proton spectrum must be accounted for self-consistently with a
non-thermal tail connecting the quasi-thermal population to the energetic
one; any change in the relativistic population must impact the thermal
population.
In contrast, the slope and normalization of test-particle
power laws can be changed without changing the temperature and density of
the shock heated gas. A description of \NL\ effects and the model used here
is given in Berezhko \& Ellison (\cite{BEapj99}), Ellison \etal\
(\cite{EBB2000}).

\section{ Modified hydrodynamic model}

\subsection{Model}

The evolution of young SNRs can be described by self-similar solutions
(Chevalier \cite{Chev82}; Nadezhin \cite{Nadezhin85}) provided that the
initial density profiles in the ejecta and in the ambient medium have
power law density distributions, respectively $\rho=g^n~r^{-n}t^{n-3}$
(with $n>5$) and $\rho=q~r^{-s}$ (with $s<3$), where $r$ is the radius,
$t$ the SNR age and $g$ and $q$ are constant. $g$ can be expressed in
terms of the explosion energy and ejected mass (Decourchelle \& Ballet
\cite{Decour94}).
The effects of cosmic ray pressure on SNR dynamics were first considered by
Chevalier (\cite{Chev83}) using a two-fluid, self-similar solution with a
thermal gas (adiabatic index $\gamma=5/3$) and a relativistic gas
($\gamma=4/3$)
having an arbitrary constant fraction of the total energy.
Here, we use Chevalier's (\cite{Chev83}) solutions for driven waves but
with boundary conditions calculated from the shock acceleration model from
Berezhko \& Ellison (1999).

As a function of shock velocity, $V_{\mathrm{s}}$, ambient density,
$\rho_0$, unshocked magnetic field, $B_0$, and ambient temperature $T_0$,
as well as the maximum particle energy, $\Emax$
\footnote{$\Emax$ is determined from the shock radius, age, and diffusive
mean free path (e.g., Ellison, Berezhko, \& Baring \cite{EBB2000}).}
and injection efficiency, $\etainjP$ (\iec the fraction of protons with
superthermal energies), the \NL\ model yields downstream values for the
density and pressures in the thermal and relativistic gases, at the forward
and reverse shocks. \alf\ wave heating in the precursor is used, reducing
the efficiency compared to adiabatic heating.
Starting with test-particle solutions at a given age, we iterate until the
modified solutions and nonlinear boundary conditions are satisfied.

Our solutions neglect cosmic ray diffusion (assuming they are spatially
coupled to the gas -- an excellent approximation for all but the
highest energy particles), and assume that both the compression ratio and
the relativistic to total pressure ratio are independent of time at
either shock, which is required for self-similarity to be valid.

Unless explicitly stated, we take $T_0= 10^4$~K and $B_0 =5$ \muG\ at both
shocks. For the SNR parameters, we take: $g= 3.4~10^9$ CGS, corresponding to
$E_{\mathrm {SN}} (M_{\odot}/M_{\mathrm{ej}})^{2/3}=0.53~10^{51}$ erg, and
$q=\rho_0=0.42$ amu/cm$^3$.

\subsection{Validity of the self-similar approach}

We can {\it a posteriori} determine the validity of the self-similar
assumption, by examining the variation versus time of the compression ratio
and relativistic to total pressure ratio, at the reverse shock (RS) and at
the forward shock (FS).
As seen in Figure ~1,
these ratios are nearly constant for injection parameters
$\etainjP \gtrsim 2~10^{-4}$ and negligible for $\etainjP~\lesssim 10^{-5}$,
so that the self-similar solutions are well justified in these cases.

For values of $\etainjP \lesssim 2\xx{-4}$, unmodified solutions can occur
at very high sonic Mach numbers $M_{\mathrm{s}}$ (i.e., early ages) which
show a rapid transition to \NL\ solutions as $M_{\mathrm{s}}$ drops (see
Berezhko \& Ellison \cite{BEapj99}). The self-similar approach is valid if
this rapid transition occurs early or late compared to the age of the SNR.

\subsection{Results on the dynamics}

Figure~2 shows the density and temperature as a function of radius for
various values of $\etainjP$. As $\etainjP$ increases, the compression ratio
becomes higher (up to 15 here instead of 4), while the shocked temperature
drops (nearly a factor of 10 between the \TP\ result and
$\etainjP=10^{-4}$).
Such trends were expected from Chevalier (\cite{Chev83}), but the actual
\NL\
effects lead to larger modifications because of the influence of particle
escape: the maximum compression ratio is no longer limited to 7, but can
rise,
in principle, to infinite values.

\begin{center}
\begin{minipage}[t]{0.47\textwidth}
\epsfxsize=0.98\textwidth \epsfbox{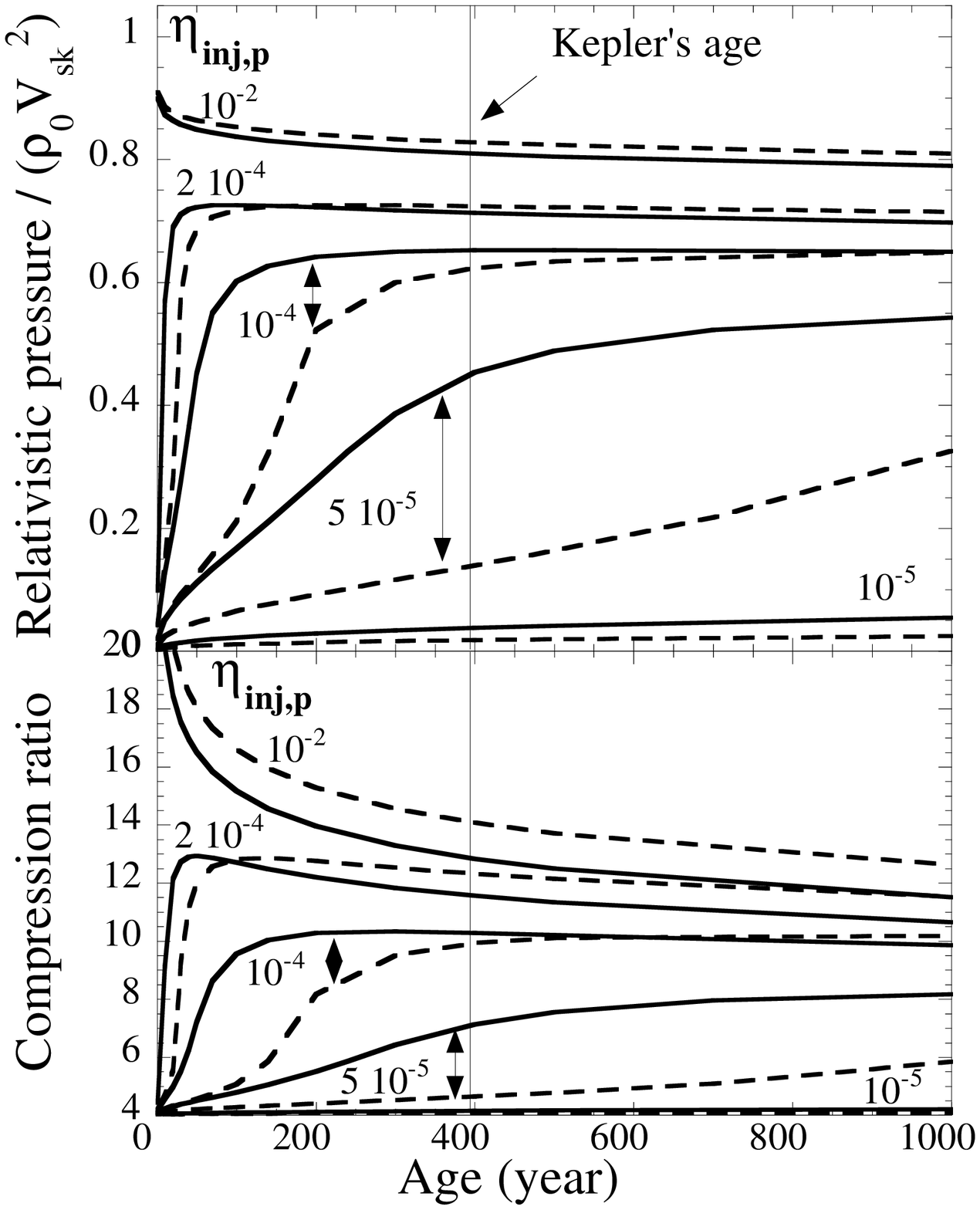}
\figcaption{As a function of remnant age for different values of $\etainjP$,
we show (top panel) the fraction of pressure in relativistic particles and
(lower panel) the compression ratio in the forward shock (dashed curves) and
in the reverse shock (solid curves).}
\end{minipage}
\end{center}

Notably, these modifications occur for $\etainjP \gtrsim 5\xx{-5}$,
considerably lower than that deduced for the Earth's bow shock and from
plasma simulations (\iec $\etainjP \sim 10^{-3}$; Ellison \etal\
\cite{EGBS93}). Table 1 gives the relative fractions of kinetic,
$E_{\mathrm{kin}}$, thermal, $E_{\mathrm{th}}$, cosmic ray,
$E_{\mathrm{cr}}$, and escaped energy, $E_{\mathrm{esc}}$, summed over the
shocked ejecta and the shocked ambient medium. An important fraction of the
energy can be carried off by escaping particles. Note that in young SNRs,
the total energy in the shocked region increases with time as more and more
ejecta is being shocked.

Cosmic ray modified dynamics lead to a thinner interaction region between
the forward and reverse shocks and much larger density and pressure
gradients.  This suggests that the growth of hydrodynamic instabilities in
the shocked material near the contact discontinuity will be faster than
otherwise.  With a forward shock closer to the contact discontinuity, this
may help these turbulent effects extend to the forward shock and beyond.

Equally important, the higher densities and lower temperatures will lead to
more rapid Coulomb equilibration than expected in a test-particle case.

\begin{center}
\begin{minipage}[t]{0.47\textwidth}
\epsfxsize=0.98\textwidth \epsfbox{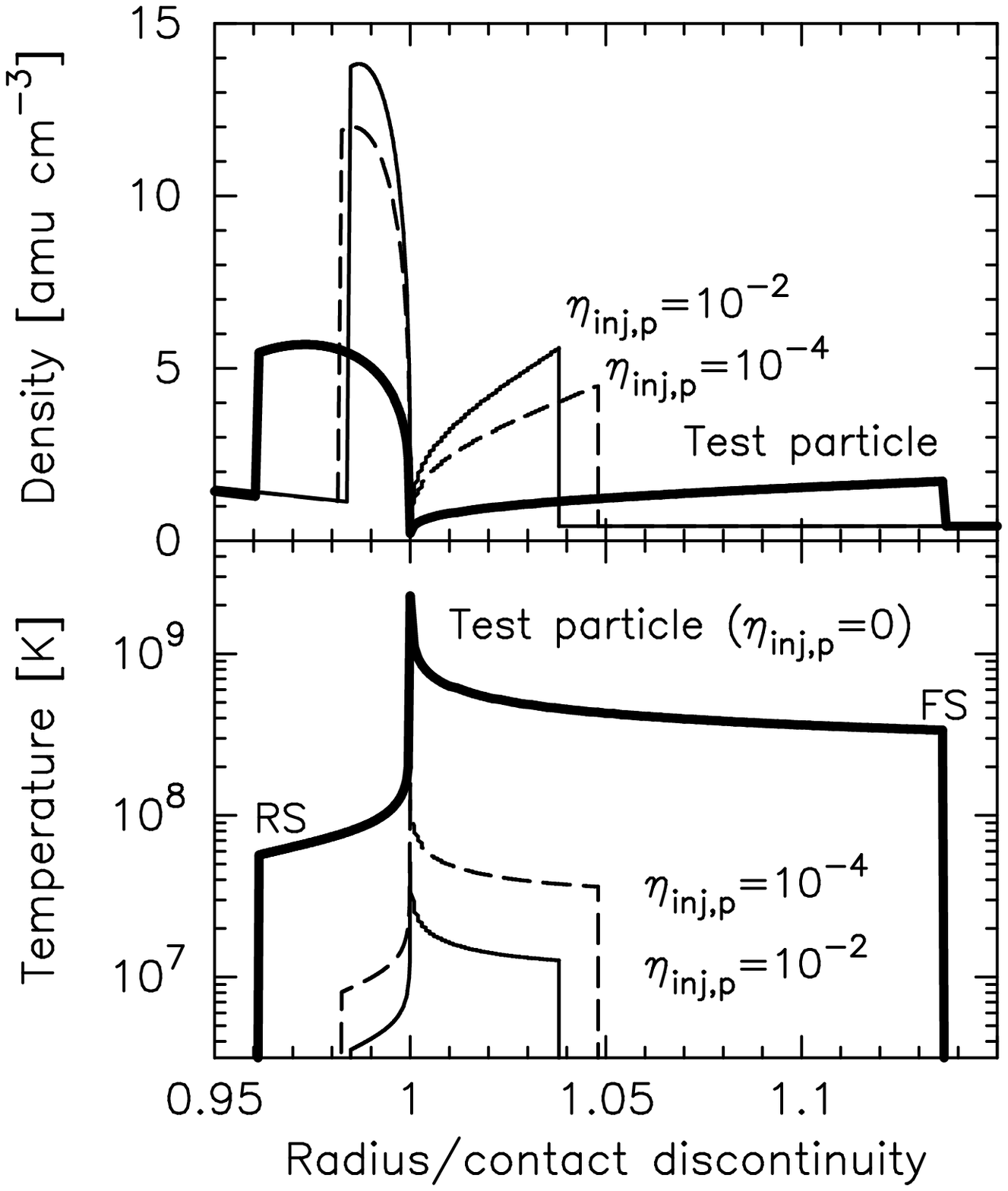}
\figcaption{Density and temperature versus radius for different values of
the injection efficiency.}
\end{minipage}
\end{center}

\subsection{Velocity versus shock temperature}

\begin{small}
\begin{minipage}[t]{83mm}
\begin{center}
{\noindent{Table 1.} \small Temperature-Velocity Relation and energy
fractions}
\begin{tabular}{lllllll}
\hline
\hline
& $\etainjP$
& $x$
& $E_{\mathrm{kin}}$
& $E_{\mathrm{th}}$
& $E_{\mathrm{cr}}$
& $E_{\mathrm{esc}}$\\
\hline
RS&$10^{-5}$ & 0.17 & 0.85 & 0.13 & 0.01 & 0.00  \\
&  $10^{-4}$ & 0.02 & 0.80 & 0.02 & 0.08 & 0.10  \\
&  $10^{-2}$ & 0.01 & 0.78 & 0.01 & 0.08 & 0.12  \\
\hline
FS&$10^{-5}$ & 0.18 & 0.48 & 0.48 & 0.03 & 0.00  \\
&  $10^{-4}$ & 0.02 & 0.42 & 0.04 & 0.26 & 0.28  \\
&  $10^{-2}$ & 0.01 & 0.41 & 0.01 & 0.24 & 0.33  \\
\hline
\end{tabular}
\end{center}
\end{minipage}
\end{small}
\vspace{.2cm}

We emphasize that when particle acceleration is efficient, the relation
between the shock velocity and the mean post-shock temperature of the
shocked gas is no longer approximated by the expression for a strong
test-particle shock, \iec $ x = [1/(\mu m_{\mathrm H})] \, (k T_{\mathrm s}
/ V_{\mathrm s}^2) \neq 3/16 $.  Here $\mu m_{\mathrm H}$ is the mean
particle mass and $T_{\mathrm s}$ is the post-shock temperature.  Table~1
shows how $x$ drops well below $3/16$ for large values of $\etainjP$, and
indicates that shock velocities derived from a measured $T_s$ may be
underestimated by a large factor. Independent measurements of $V_s$ and
$T_s$ provide strong constraints on the amount of energy going into
relativistic particles.
This was shown using {\it Chandra} observations of E0102.2-7219 SNR, by
Hughes, Rakowski, \& Decourchelle (2000), who concluded that the low
observed post-shock electron temperature combined with the high observed
shock velocity could only be understood if a non-negligible fraction
of the shock ram energy went into cosmic rays.

\section{Interpretation of X-ray thermal spectra}

\vspace{.2cm}
\begin{small}
\begin{table*}
\begin{center}
{\noindent{Table 2.} Best-fit derived quantities for different values
of the injection parameter at both shocks.}
\begin{tabular}{lllllllllll}
\hline
\hline
$\eta_{\mathrm{inj,p FS}}$ & $\eta_{\mathrm{inj,p RS}}$
& $kT_{\mathrm {FS}}$     & $kT_{\mathrm {RS}}$
& $r_{\mathrm {FS}}$     & $r_{\mathrm {RS}}$
& $V_{\mathrm {FS}}$     & $V_{\mathrm {RS}}$         & $\rho_0$    & $D$
& $E_{\mathrm {SN}} (M_{\odot}/M_{\mathrm{ej}})^{2/3}$ \\
           &                    &  keV               & keV
           &                    &
           & km/s               & km/s               & amu/cm$^{3}$ & kpc
           & $10^{51}$ erg\\
\hline
 0          &  0          & 17  &  3 & 4  &  4 &  4000 & 1700 & 0.67 & 5  &
0.3\\
$2~10^{-4}$ & $2~10^{-4}$ & 14  &  3 & 14 & 13 & 12200 & 5800 & 0.23 & 15 &
9\\
$  10^{-3}$ & $  10^{-3}$ & 11  &  3 & 17 & 16 & 15800 & 7600 & 0.18 & 20 &
18\\
$  10^{-2}$ & $  10^{-2}$ & 11  &  3 & 18 & 17 & 18200 & 8800 & 0.17 & 23 &
30\\
\hline
$10^{-3}$ &  0          & 1.5 &  5 & 13 &  4 &  4570 & 2110 & 0.57 & 5  &
0.5\\
\hline
\end{tabular}
\end{center}
\end{table*}
\end{small}

The increase in shocked density and decrease in shocked temperature
from \NL\ effects, strongly impact the thermal X-ray emission.
Compared to test-particle models, we predict that regions in young
SNRs undergoing efficient acceleration will have a lower temperature, more
intense thermal X-ray emission (due to the larger emission integral
$\int{n_e^2~dV}$ and lower temperature), and a faster ionization timescale.
In older remnants, the temperature should drop below X-ray
emitting values earlier.

We have coupled our modified hydrodynamics model to the calculation of the
non-equilibrium ionization and thermal X-ray emission. The ionization and
recombination rates have been taken from Arnaud \& Rothenflug
(\cite{Arnaud85}) and Arnaud \& Raymond (\cite{Arnaud92}). The X-ray
emission code is from Mewe, Gronenschild, \& Van den Oord (\cite{Mewe85})
and Mewe, Lemen, \& Van den Oord (\cite{Mewe86}). The abundances in the
interstellar medium are taken to be solar (Meyer \cite{Meyer85}). We assume
full equipartition between the electrons and
ions ($T_e = T_i$) and neglect any contribution from the superthermal
electrons to the ionization or line excitation. For efficient shock
acceleration, this may be important and we are investigating these effects
(Porquet et al. 2000).

\begin{center}
\begin{minipage}[t]{0.47\textwidth}
\epsfxsize=0.98\textwidth \epsfbox{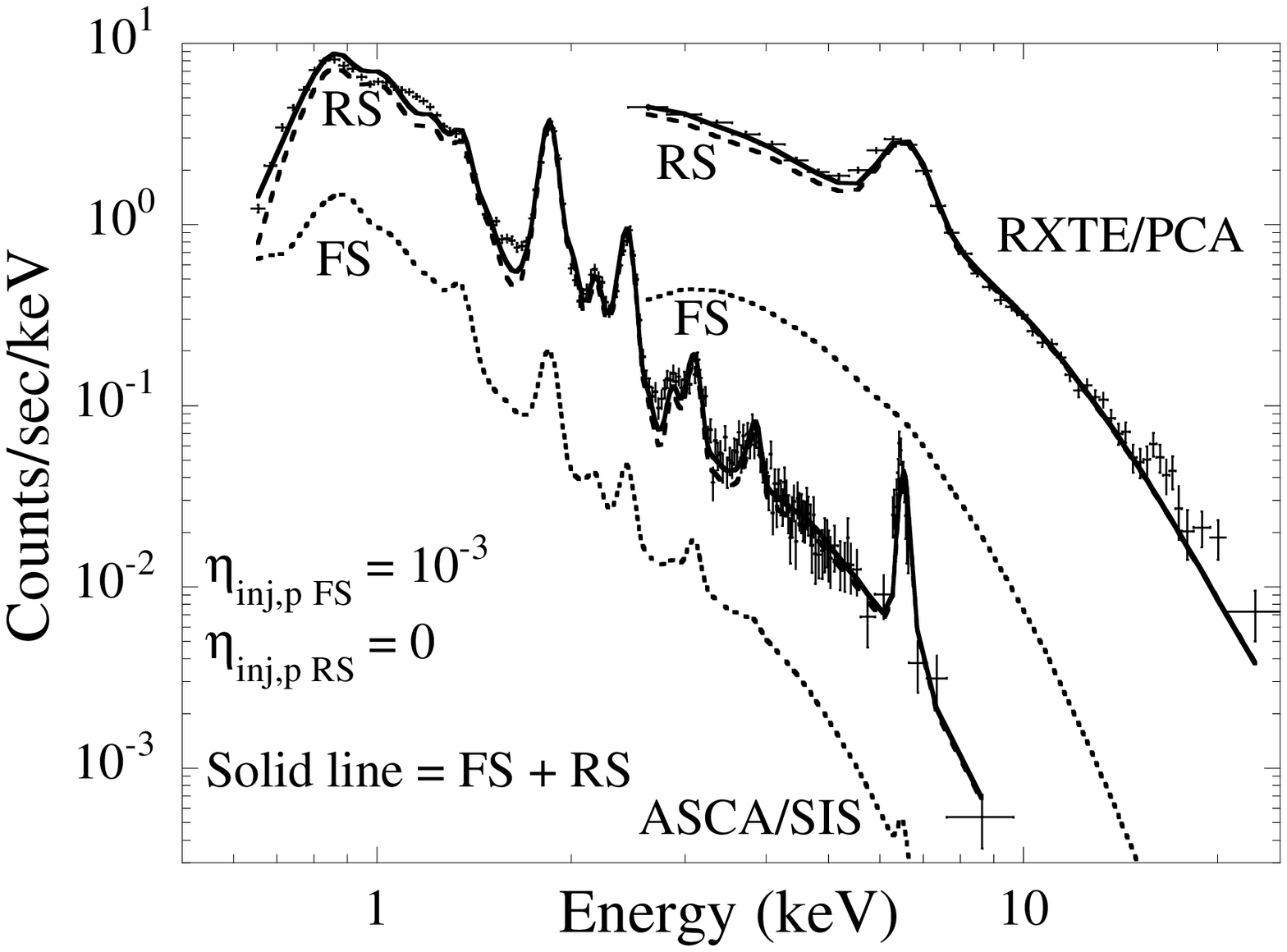}
\figcaption{Best-fit model to the spectra of Kepler's SNR. The
contribution from the shocked ejecta (RS) is shown with dashed lines,
the shocked ambient matter (FS) is shown with dotted lines, and the sum is
shown with solid lines.}
\end{minipage}
\end{center}

As an indication of how the interpretation of X-ray data depends on
acceleration efficiency, we model the X-ray observations of Kepler's SNR
obtained over a broad energy range with {\it ASCA} (Kinugasa \& Tsunemi
1999) and {\it RXTE} (Decourchelle \& Petre 1999; Decourchelle et
al. 2000). For several values of $\etainjP$, we adjust the SNR parameters
to fit the observed spectrum, which is dominated by the emission from the
shocked ejecta as shown in Figure~3. All values of $\etainjP$ yield similar
values for the reduced chi-squared ($\sim 4$), the ionization timescale,
the reverse shock temperature $\sim 3$ keV, the hydrogen column density
$\sim 4~10^{21}$~cm$^{-2}$, and oversolar abundances of heavy elements in
the shocked ejecta (O, Ne, Mg, Si, S, Ar, Ca, and Fe).
The derived values of forward and reverse shock compression ratios
($r_{\mathrm{FS}}$ and $r_{\mathrm{RS}}$), FS and RS speeds
($V_{\mathrm{FS}}$
and $V_{\mathrm{RS}}$), ambient density, distance, and the energy-mass
relation are, however, very different as shown in Table~2.

To get a temperature high enough to produce the iron K-line and the
high energy continuum, all \NL\ fits require larger shock velocities
and lower ambient densities. These imply more kinetic energy in the
ejecta and a larger inferred distance, $D$, than the \TP\ case.  To
obtain a distance consistent with observations (4.8 $\pm$ 1.4 kpc,
Reynoso and Goss \cite{Reynoso99}), a low efficiency at the reverse shock
is required.  This can be done in three ways: with $B_0 \gtrsim 25$\muG\
at the reverse shock large enough to dampen acceleration, with $B_0
\lesssim 0.1$ \muG\ and $\eta_{\mathrm{inj,p RS}} \lesssim 10^{-4}$ to
allow high Mach number, unmodified solutions, or with no reverse shock
acceleration ($\eta_{\mathrm{inj,p RS}} \simeq 0$). A low magnetic field
is expected from the expansion of the ejecta, but magnetic field
amplification could increase it. In all cases, a good fit is obtained
for efficient acceleration at the forward shock and test-particle
conditions at the reverse shock. Figure 3 and Table 2 (last row) show
the case with $\eta_{\mathrm{inj,p RS}} = 0$.

Spatially resolved spectroscopy is required to confirm both a low
efficiency of the acceleration at the reverse shock and our prediction
of efficient acceleration at the forward shock, which is expected for
standard values of the upstream density and magnetic field. The
arbitrary power law component required to fit the broad-band X-ray
continuum (Decourchelle \& Petre \cite{Decour99}) is likely related to
the forward shock.  In future work, this component will be replaced by
including self-consistently the nonthermal continuum, and using
additional constraints on the acceleration parameters from the radio
data.

\section{Conclusion}

Nonlinear effects from efficient particle acceleration strongly
influence the X-ray emission in SNRs, and may serve as a tracer
of cosmic ray acceleration.
Here, we have focused on the consequences of shock acceleration on the
thermal X-ray emission and have presented the first \NL, non-equilibrium
ionization modeling of the X-ray emission from a specific SNR.  While our
self-similar approach has limitations which must be carefully considered,
we believe it contains the essential physics and allows the exploration of
a large range of parameters.
We find that typical source parameters of young SNRs should result in
\NL\ acceleration at the forward shock, but yield near test-particle
conditions at the reverse shock if the magnetic field is as low as expected
from the ejecta expansion.

When cosmic ray production is efficient, shock {\it heating} is linked to
particle {\it acceleration} and thus to the broad-band photon emission from
energetic electrons and ions.  The high energy electrons produce both radio
and X-ray synchrotron and this will be included in future work.

In addition to a possible nonthermal synchrotron component, regions
undergoing strong acceleration should exhibit stronger and lower
temperature thermal emission, while regions with little acceleration should
be fainter and at a higher temperature.

Also, other complications (e.g., the variation of the ambient magnetic
field orientation (Reynolds \cite{Reynolds98}), shock interactions with
clouds and dense knots, etc.)  influence the full modeling of SNRs. The
spatially resolved spectra from the {\it Chandra} and {\it XMM-Newton}
satellites should greatly aid the investigation of these and other effects.

\acknowledgments
 We thank E. Berezhko, L. Drury, J. Hughes, J.P. Meyer and E. Parizot for
fruitful discussions as well as the referee for helpful comments.

\end{document}